
\input amstex
\documentstyle{amsppt}
\NoRunningHeads

\hsize 6.25truein
\hfuzz=25pt
\vsize 8.5truein

\TagsOnRight

\def\oh#1{\mathaccent'27#1} 

\topmatter
Math. Models and Methods in Appl. Sci., 9, N2, (1999), 325-335.
\title  A numerical method for solving nonlinear problems  \endtitle
\author  A.G. Ramm      \endauthor
\affil  Department of Mathematics, Kansas State University, 
        Manhattan, KS  66506-2602, USA \\
        and IMAG-UJF, France
\endaffil
\address      \endaddress
\email ramm\@math.ksu.edu \endemail
\thanks
The author thanks Dr. V. Protopopescu for useful comments 
\endthanks


\endtopmatter

\vglue .1in

\document

\subhead 1. Introduction,  basic idea and results\endsubhead


Suppose
$$ A(u)=f, $$  \vskip-.35in $$\tag1.1 $$

$$ A(0)=0, $$ \vskip-.35in $$\tag1.2 $$

where $A(u)$ is a nonlinear
 operator on a Banach space $X$, which has the G- 
derivative:
$$ A'(u)w=\lim_{t\to 0}\ \frac{A(u+tw)-A(u)}{t},
   \qquad \forall w\in D(A). \tag1.3 $$
Here $D(A)$ is the domain of definition of the operator $A$,
which is assumed to be a set,  such that if $u$
and $w$ belong to it, then for all sufficiently small $t\in \Bbb R$,
the element $u+tw$ belongs to it. The operator
$A'(u)$ is a linear, maybe  unbounded, densely defined
 operator on $X$. Even when $A'(u)$ is unbounded, it may be that 
the following estimate holds: 
$$\Vert A'(u)-A'(v)\Vert \to 0 \text { as }\Vert u-v\Vert \to 0.$$
This is usually the case when the unbounded part of  $A'(u)$ 
is an operator independent of $u$, see Example 2 in section 2 below.

In this paper we develop an idea from [1] to write (1.1) as an equation:
$$ L(u)u=f, \quad A(u)=L(u)u,\tag1.4 $$
where $L(u)$ is a {\it linear} operator on $X$ which depends
{\it nonlinearly} on $u(x)$ as on a parameter. 
We assume that 
$L(u)$ has bounded inverse, that is $L^{-1}(u)$ is a bounded
linear operator defined on all of $X$.

We study an approach described in [1] for solving equation (1.1)
written in the form (1.4). This approach differs from the
traditional ones (such as Newton's method and its modifications,
and other methods, based on {\it local} linearizations of
nonlinear operator $A(u)$ (see [2] for a detailed study of such methods).
The difference is in the {\it global}
 nature of equation (1.4)
with linear operator $L(u)$ which depends nonlinearly on $u$.

In [1] the goal was to construct an analog of Green's function for nonlinear 
systems and to suggest numerical schemes, which would be 
based on the schemes for solving linear equations. The hope is that
such schemes may be more efficient and more stable computationally.
In [1] an iterative scheme (1.11)-(1.12) with $u_0=f$ was
suggested.
 Analysis of the convergence of this iterative scheme 
 was not given in [1] but numerical 
examples in [1] are encouraging. Here we give such an analysis.

In Theorem 1.1 we give sufficient conditions for convergence of
an iterative process (1.11)-(1.12) (see below) for solving
equation (1.1). These conditions are formulated in terms of the
operator $L(u)$ defined in (1.6) below.

In Theorem 1.2 sufficient conditions are given in terms of the 
G-derivative of $A(u)$ for the operator $L(u)$ to satisfy the conditions
of Theorem 1.1.

In section 2 examples are briefly considered.

For linear equation (1.4) one can use various methods for 
constructing $L^{-1}(u)$. One writes (1.4) as
$$u=L^{-1}(u) f, \tag1.5 $$
assuming that $L(u)$ is boundedly invertible ( sufficient
conditions for this are given in Theorem 1.2)  and uses an 
iterative scheme to solve  equation (1.5).

Let us first prove that
$$L(u)=\int^1_0 A'(tu)\,dt,\tag1.6 $$
and that
$$ A(u)=\int^1_0A'(tu)\,dt\,u=L(u)u. \tag1.7 $$
The following lemma is an immediate consequence of the definition
of the G-derivative and its proof
is included for convenience of the reader.
\proclaim{Lemma 1.1} Formula (1.7) holds. \endproclaim

\demo{Proof of Lemma 1.1} Note that
$$ \int^1_0\frac{d}{dt} A(tu)\,dt= 
   A(u)-A(0)=A(u),\tag1.8 $$
where the assumption (1.2) was used.

On the other hand,
$$ \frac{d}{dt}\ A(tu) = \lim_{s\to 0} 
   \frac{A\bigl( (t+s)u\bigr)-A(tu)}{s} = A'(tu)u, \tag1.9 $$
by the definition of the G-derivative. It follows 
from (1.8), (1.9), and the linearity of $A'(u)$ as an operator 
on $X$, that
$$ \int^1_0A'(tu)\,dt\,u=A(u),\tag1.10 $$
as claimed.

Lemma 1.1 is proved. \qed
\enddemo

Now the linear operator $L(u)$ on $X$ (depending on $u$ 
nonlinearly) is well defined by formula (1.6), and equation (1.4)
is well understood as an equation for $u$ with linear operator 
$L(u)$, which depends nonlinearly on $u$.

If $L(u)$ is boundedly invertible, that is, $L^{-1}(u)$ is a 
bounded linear operator defined on all of $X$, then (1.4) is
equivalent to (1.5). Below we use the boundedness of $L^{-1}(u)$ 
not for all $u\in X$ but only in a ball $B_R:=\Vert u\Vert \leq R$,
where $R>0$ is some number  which  will be specified in the proof
of Theorem 1.1.

Let us write an iterative process for solving  equation (1.5):
$$ u_{n+1}= L^{-1}(u_n) f,\tag1.11 $$
$$ u_0=u_0,\tag1.12 $$
where $u_0$ is an initial approximation which is arbitrary
at this moment. In practice it should be chosen
taking into account all the a priori information available about the solution.
This  may improve the rate of convergence of the scheme (1.11)-(1.12).

Let us now analyze the above iterative process.

We have
$$ \Vert u_{n+1}-u_n\Vert  = \Vert L^{-1}(u_n)-L^{-1}(u_{n-1})\Vert  
  \ \Vert f\Vert \leq q\Vert  u_n-u_{n-1}\Vert \ \Vert f\Vert . 
  \tag1.13 $$
Here we have assumed that:
$$ \Vert L^{-1}(u)-L^{-1}(v)\Vert  \leq q \Vert u-v\Vert , \tag1.14 $$
for $u,v\in B_R$, where $B_R$ is a ball 
$\Vert u\Vert \leq R$, $R>0$ is a certain number, and $q>0$ is a number 
independent of $u,v\in B_R$. 

Assume now that 
$$ q \Vert f\Vert \leq Q<1 .\tag1.15 $$
Then 
$$\eqalign{
   \Vert u_1-u_0\Vert  & \leq \Vert L^{-1}(u_0) f-u_0\Vert , \cr
   \Vert u_2-u_1\Vert  & \leq \Vert L^{-1}(u_1)-L^{-1}(u_0)\Vert 
     \ \Vert f\Vert  \leq q \Vert f\Vert \ \Vert u_1-u_0\Vert   
     \leq Q \Vert u_1-u_0\Vert,  \cr
   \Vert u_{n+1}-u_n\Vert  & \leq \Vert L^{-1}(u_n)-L^{-1}(u_{n-1})\Vert  
     \ \Vert f\Vert  \leq Q\Vert u_n-u_{n-1}\Vert  \cr
    & \leq Q^2\Vert u_{n-1}-u_{n-2}\Vert  
    \leq \cdots \leq Q^n \Vert u_1-u_0\Vert . } $$
Thus
$$ \Vert u_{n+1}\Vert  
   = \bigg\Vert u_0+\sum^{n+1}_{j=1} (u_j-u_{j-1})\bigg\Vert 
   \leq \Vert u_0\Vert  + \Vert u_1-u_0\Vert \ \frac{1-Q^{n+1}}{1-Q}
   \leq \Vert u_0\Vert  + \frac{\Vert u_1-u_0\Vert }{1-Q}. \tag1.16 $$

It is sufficient to assume that conditions (1.14) -(1.15) hold in a ball 
$B_R$ such that $B_R\supset S$, where
$$ S:=\{u:\Vert u\Vert  
   \leq \Vert u_0\Vert  + (1-Q)^{-1} \Vert u_1-u_0\Vert \}. \tag1.17 $$
If $B_R\supset S$, then
all the members of the sequence $u_n$ belong to $S$,
and the sequence $u_n$ converges to the element
$$u:=u_0+ \sum^\infty_{j=1} (u_j-u_{j-1}),\qquad u\in S. \tag1.18 $$

Passing to the limit in (1.11) and using condition (1.14), which implies
the continuity
(in the operator norm) of $L^{-1}(u)$ with respect to $u$, one gets
$$ u=L^{-1}(u)f, \tag1.19 $$
or
$$ A(u)=L(u)u=f. \tag1.20 $$

Let us formulate the result.
   
\proclaim{Theorem 1.1} Assume that $A(u)$ has the G-derivative 
$A'(u)$ which is a linear, possibly unbounded, densely defined
 operator on $X$,  such that for the operator
 $L(u)$, defined in equation (1.6), the conditions  
(1.14) and (1.15) hold for all $u$ and $v$ in $S$, 
where $S$ is defined in (1.17). 
Then equation (1.1) is uniquely solvable in $S$ and its solution can 
be obtained by the iterative process (1.11)-(1.12).
\endproclaim

\demo{Remark 1.1} One could choose $u_0=f$ in (1.12). However, 
in practice there may be some information available which allows one 
to choose a better initial approximation $u_0(x)$ to the unknown 
solution $u$. Note that our conditions (1.14)-(1.15) are the usual
conditions for the applicability of the contraction mapping principle
to the  equation $u=L^{-1}(u)f$. 

One could formulate a condition, which implies (1.15), in terms
of the bounds on the derivative of $L^{-1}(u)$ with respect to $u$.
Namely $L^{-1}(u)_{u}=-L^{-1}(u)L(u)_u L^{-1}(u),$ where
the subscript $u$ denotes the G-derivative. Therefore
(1.14) holds with $q=\sup_u ||L^{-1}(u)L(u)_u L^{-1}(u)||.$
We do not go into further detail since conditions (1.14), (1.15)
are simple and can be verified in some examples (see
for instance example 2 in section 2).

Note that an application of the contraction mapping principle
directly to the equation (1.1) is not possible in many practical
problems in which the operator $A(u)$ is unbounded. Therefore
equation (1.5) is much more suitable for an application
  of the contraction mapping principle in these problems.
\enddemo

\demo{Remark 1.2} The inequality, 
which implies that $S\subset B_R$,  is of the form:
$$\Vert u_0\Vert  + \frac{\Vert L^{-1}(u_0)f 
  -u_0\Vert }{1-q\Vert f\Vert }\leq R. \tag1.21 $$
Solving it for $q$, one gets the following inequality:
$$0<q<\Vert f\Vert ^{-1} 
   \left( 1-\frac{\Vert L^{-1}(u_0)f-u_0\Vert }{R-\Vert u_0\Vert } 
   \right).\tag1.22 $$
For a given $f$ ($f\not= 0$ since otherwise $u=0$ by the assumption 
(1.2) and the local injectivity of $A(u)$ near $u=0$, which 
follows from the assumed bounded invertibility of $L(0)$ ), one can fix
an arbitrary $u_0$ and then choose $R$,  
$R>\Vert u_0\Vert $, sufficiently large, so that
$$ \frac{\Vert L^{-1}(u_0)f-u_0\Vert }{R-\Vert u_0\Vert } <1,\tag1.23 $$
and
$$ 0<Q<1, \quad Q:= 1- \frac{\Vert L^{-1}(u_0)f-u_0\Vert }{R-\Vert u_0\Vert } 
   . \tag1.24 $$
Then there exists $q>0$ which satisfies (1.22) 
and therefore (1.15) holds.

 In particular, if
 $$u_0=0, \quad \text { and } R>\Vert L^{-1}(0)f\Vert, 
$$ 
then equation (1.1) is solvable in the ball 
$B_R$ provided that (1.14) holds and
$$0<q<\Vert f\Vert ^{-1}\left(1-\frac{\Vert L^{-1}(0)f\Vert }{R}\right). 
$$
Assumptions  (1.14)-(1.15) imply the unique solvability of equation (1.1) 
in the ball $B_R$. This follows from Theorem 1.1, but we give an independent simple proof of this claim.

 Indeed, suppose
$$ A(u)=L(u)u=f, \quad A(v)=L(v)v=f, \qquad u,v\in B_R.\tag1.26 $$
Then
$$ L(u)u=L(v)v, \quad u=\left(L^{-1}(u)-L^{-1}(v)\right)f+v, $$
so, using (1.14) and (1.15), one gets
$$ \Vert u-v\Vert  \leq \Vert L^{-1}(u)-L^{-1}(v)\Vert  
   \ \Vert f\Vert  \leq Q \ \Vert u-v\Vert <\Vert u-v\Vert .
   \tag1.27 $$
It follows from (1.27) that $u=v$. This proves uniqueness of the 
solution to (1.1) in the ball $B_R$. Existence  of the solution follows from Theorem 1.1.
\enddemo

Now we give sufficient 
conditions for the existence of the bounded  inverse operator $L^{-1}(u)$. 

We start with

\proclaim{Lemma 1.2} Assume that:
\vskip.1in
i) $\left\Vert \left[A'(0)\right]^{-1}\right\Vert  \leq p,$
\vskip.1in
ii) $\sup_{0\leq t\leq 1} \Vert A'(tu)-A'(0)\Vert \leq s$, 
    $\forall u\in B_R:= \{u:\Vert u\Vert \leq R\}$, 
    where $R>0$ is some number, 

and
\vskip.1in
iii) $ps<1.$  

\noindent Then 
$$ \sup_{0\leq t\leq 1} \left\Vert  \big[A'(tu)\big]^{-1}\right\Vert  
   \leq\frac{p}{1-ps}, \qquad\forall u\in B_R.\tag1.28$$
\endproclaim

\demo{Proof} 
First note that in (1.28) existence of the bounded linear operator
$\big[A'(tu)\big]^{-1}$ is also claimed. One has
$$ A'(tu)=A'(0)+A'(tu)-A'(0)=A'(0) \left[ I+\big(A'(0)\big)^{-1}
   \big( A'(tu)-A'(0)\big) \right]:=A'(0)(I+B). $$
Therefore
$$ \big[A'(tu)\big]^{-1} =(I+B)^{-1} \big[A'(0)\big]^{-1}, 
   \tag1.29 $$
where both inverse operators on the right-hand side of 
equation (1.29) exist, and by i) and ii), one has:
$$ \Vert B\Vert \leq ps<1. $$
Therefore:
$$ \Vert (I+B)^{-1}\Vert \leq \frac{1}{1-ps}. \tag1.30 $$
From (1.29), (1.30) and the assumption i), the desired conclusion 
(1.28) follows. Lemma 1.2 is proved.
\qed
\enddemo

Now we want to prove that, under the assumptions of lemma 1.2, the operator 
$L^{-1}(u)$ does exist and is bounded.
 
Since 
$$L(u)=\int^1_0A'(tu)\,dt,$$
 one has
$$ L(u)=A'(0) 
  \left[ I+\big(A'(0)\big)^{-1} 
  \int^1_0\big(A'(tu)-A'(0)\big)\,dt \right]. \tag1.31 $$
Using i) and ii) of Lemma 1.2, and denoting $[A'(0)]^{-1}:=T$, 
one gets
$$ L^{-1}(u)= [I+TM]^{-1}\ T, \tag1.32 $$
where
$$ M:= \int^1_0 \big[ A'(tu)-A'(u) \big]\,dt,
   \quad \Vert M\Vert \leq s, \quad \Vert TM\Vert \leq ps\leq 1.\tag1.33 $$
Therefore
$$ \Vert (I+TM)^{-1}\Vert \leq \frac{1}{1-ps}.\tag1.34 $$

We have proved the following:

\proclaim{Theorem 1.2} Assume i), ii) and iii) of Lemma 1.2.
Then $L^{-1}(u)$ exists for all $u\in B_R$ and
$$ \Vert L^{-1}(u)\Vert \leq \frac{p}{1-ps}. \tag1.35 $$
\endproclaim

\subhead 2. Examples \endsubhead

\vskip .1in

\subsubhead Example 1 \endsubsubhead

Let $X=C(D)$, the usual Banach space of continuous in the
 domain $D$ functions, 
 $K(x,y,u)$ and $K_u(x,y,u)$ be continuous functions on 
$D\times D\times{\Bbb R}$, and
$$ A(u):=u+B(u):=u+\int_D K\left(x,y,u(y)\right)\,dy, \quad K(x,y,0)=0. $$
We have
$$ A'(u)w=w(x)+\int_D \frac{\partial K}{\partial u} 
   \left( x,y,u(y)\right)w(y)\,dy, $$
and
$$ L(u)w=w(x)+\int_D\int^1_0 \frac{\partial K}{\partial u}
   \left(x,y,tu(y)\right)\,dt
   \, w(y)\,dy=w(x)+\int_D\frac{K\left(x,y,u(y)\right)}{u(y)}w(y)\,dy. $$
This $L(u)$ is a continuous linear operator on $C(D)$ which
depends on $u(y)$ nonlinearly.

Equation
$$ A(u)=f \tag2.1 $$
is equivalent to 
$$ L(u)u=f, \tag2.2 $$
or, if $L^{-1}(u)$ is a bounded linear operator, to
$$ u=L^{-1}(u)f. \tag2.3 $$
An iterative scheme
$$ u_{n+1}=L^{-1}(u_n)f,\ \text{or}\ L(u_n)u_{n+1}=f,\,\, u_0=u_0, \tag2.4 $$
can be studied and, under suitable assumptions on $L(u)$, 
this scheme converges.

Let us briefly outline a proof of the boundedness of $L^{-1}(u)$
in this example. Assume that $K(x,y,u)=k(x,y)g(u)$, where
$k(x,y)$ is a continuous selfadjoint kernel, and $g(u)$ is a continuous
function such that $\frac {g(u)} u>0$. Assume that $u(x)>0$ is
a continuous  function in the closure of $D$.
 Then $L(u)$ is a Fredholm-type operator with index zero, and
its null-space is trivial, as one can easily check using the above
assumptions. By Fredholm's alternative, $L^{-1}(u)$ is bounded.
If the selfadjoint  kernel $k(x,y)$ has positive eigenvalues, and if
$g'(u)>m>0$, then equation (2.1) has no more than one solution.
Indeed, if $A(u)-A(v)=0$, then 
$$(u-v, g(u)-g(v)) +(k[g(u)-g(v)],g(u)-g(v))=0
$$
where the parentheses denote the inner product in $L^2(D)$, 
$kw:=\int_Dk(x,y)w(y)dy$, and
our assumptions imply that both terms in the above equations are nonnegative.
Therefore they both vanish, and this implies that $u=v$ as claimed.
 
\vskip.1in
\subsubhead Example 2 \endsubsubhead

Let
$$ A(u):=-\Delta u+g(x,u)=f(x),\ \text{in}\ D,\quad u\big|_S=0,
   \quad g(x,0)=0, \tag2.5 $$
where $D\subset R^n$ is a bounded domain with sufficiently smooth
boundary $S$, $f(x)$ is a given function, $g(x,u)$ is continuous 
with respect to $x\in D$ and $C^1$ with respect to $u\in R$. 
As $X$ we take $H=L^2(D)$. 
One has
$$ A'(u)w=-\Delta w+g_u'(x,u)w, $$
and
$$ L(u)w=-\Delta w +\frac{g\left(x,u(x)\right)w}{u(x)}. \tag2.5 $$
The operator $L(u)$ here is unbounded and his
unbounded part, $-\Delta$, does not depend on $u(x)$. Under 
suitable assumptions 
the operator $L(u)$ is boundedly invertible and one can estimate 
$L^{-1}(u)$. For example, suppose $g(x,u)>0$ and  
$\frac{g\left(x,u(x)\right)}{u(x)}=a+q(x)$, where $a=\text{const}>0$, and 
$q(x)$ is a nonnegative function which depends on $u(x)$. 
In this case
$$ L^{-1}(u)=\left(-\Delta+a+q(x)\right)^{-1},
   \quad \Vert L^{-1}(u)\Vert \leq\frac{1}{a}. $$
Here the operator $-\Delta+a+q(x)$ is defined in $L^2(D)$ 
by the Dirichlet boundary condition, or by its domain 
$H^2(D)\cap \oh H^1(D)$, where $H^m(D)$ are 
the Sobolev spaces.

One can get the following estimate:
$$\eqalign{
  \Vert L^{-1}(u)-L^{-1}(v)\Vert  &\leq \Vert L^{-1}(u)\Vert 
   \ \Vert L(u)-L(v)\Vert  
   \ \Vert L^{-1}(v)\Vert  \cr
   & \leq \frac{1}{a^2} \bigg\Vert \frac{g(u)}{u} 
     -\frac{g(v)}{v}\bigg\Vert  
     \leq \frac{c(g)}{a^2} \Vert u-v\Vert .} \tag2.7 $$
In (2.7) we have assumed that
$$ c(g)=\max_{u\in R}
   \left|\left(\frac{g(u)}{u}\right)_u'\right| <\infty, \tag2.8 $$
and have used
the following well-known identity:
$$B_1^{-1}-B_2^{-1}= -B_1^{-1}(B_1-B_2)B_2^{-1}, \tag2.9 $$
with $B_1=L(u)$, $B_2=L(v)$.
Thus, the conditions (1.14)-(1.15), sufficient for the convergence
of the iterative process (2.4), hold if $c(g) ||f|| a^{-2}<1$. 
\vskip.1in

\subsubhead Example 3 \endsubsubhead

Let $D\subset R^n$ be a bounded domain with a sufficiently smooth
boundary $S$. Consider the problem
$$\frac{\partial u}{\partial t}-\nabla\cdot
  \left[a(u)\nabla u\right] =f(x,t)
  \ \text{in}\ D\times [0,T],\tag2.10 $$
$$ u\big|_{t=0}=u_0(x), \tag2.11 $$
$$ u\big|_S=0. \tag2.12 $$
Write this problem as
$$ A(u):= u(x,t) -\int^t_0B(u)\,d\tau
  =u_0(x)+\int^t_0 f(x,t)\,d\tau \tag2.13 $$
with $u=u(x,t)\in C^1\left(\oh H^2(D), [0,T]\right)$, 
and
$$B(u):=\nabla\cdot[a(u)\nabla u]. \tag2.14 $$
Assume that
$$ a\in C^2(R),\quad 0<c\leq a\leq m, 
   \quad |a'|,\ |a''|\leq m, \tag2.15 $$
where $c$ and $m$ are some positive constants and the primes
denote derivatives with respect to $u$.

One checks that
$$A'(u)w=w- \int^t_0 B'(u)w\,d\tau ,\tag2.16 $$
where
$$ B'(u)w =\nabla\cdot [ a'(u)\nabla w ] 
   +a'(u)(\nabla^2 u)w +a''(u)(\nabla u)^2w 
   +a'(u)\nabla u\cdot \nabla w. \tag2.17 $$

Denote
$$ \gamma(u):=\int^u_0 a(v)\,dv. \tag2.18 $$
Now one checks that
$$\eqalign{ 
 & \int^1_0 B'(\alpha u)\,d\alpha\, w=\nabla\cdot
   \left[\frac{a(u)-a(0)}{u}\nabla w\right]
   +\left(\frac{a(u)}{u}-\frac{\gamma(u)}{u^2}\right)
   (\Delta u )\,w \cr
 & \quad +\left(\frac{a(u)}{u}-\frac{\gamma(u)}{u^2}\right)
   \nabla u\cdot\nabla w +
   \left(\frac{a'(u)}{u}-\frac{2a(u)}{u^2}+\frac{2\gamma(u)}{u^3}\right)
   (\nabla u)^2 w.} $$ 
\vskip-.35in
$$ \tag2.19 $$

Therefore,
$$\eqalign{ 
  L(u)w
 & =\int^1_0 A'(\alpha u)\,d\alpha\,w=w-\int^t_0
   \int^1_0 B'(\alpha u)\,d\alpha\,w\,d\tau  \cr
 & =w-\int^t_0
   \left\{ 
      \nabla\cdot\left[ \frac{a(u)-a(0)}{u}\nabla w \right]
      + \left( \frac{a(u)}{u}-\frac{\gamma(u)}{u^2} \right) 
      (\Delta u)\,w \right. \cr
 & + \left.\left( \frac{a(u)}{u}-\frac{\gamma(u)}{u^2} \right) 
      \nabla u\cdot\nabla w+
      \left( \frac{a'(u)}{u}-\frac{2a(u)}{u^2}+\frac{2\gamma(u)}{u^3} \right)
      (\nabla u)^2 w  \right\} d\tau  \cr}
 $$
\vskip-.35in
$$\tag2.20 $$
and this equation defines a linear unbounded operator $L(u)$,
which depends on $u=u(x,t)$ nonlinearly as on a parameter.

It seems at first sight that as $u\to 0$ one has difficulties
in the definition (2.20), since 
$u$ is in the denominator. However,  in fact there are no difficulties:
one checks using L'Hospital rule,
for example, that the functions
$$\frac{a(u)}{u}-\frac{\gamma(u)}{u^2}=\frac{ua(u)-\gamma(u)}{u^2}
  \to \frac{a'(0)}{2} \tag2.21 $$
and
$$ \frac{a'(u)u^2-2ua(u)+2\gamma(u)}{u^3}
  \underset{u\to 0}\to\rightarrow \frac{a''(0)}{3} \tag2.22 $$
are bounded as $u\to 0$.

\vfill

\enddocument